\documentclass[12pt]{iopart}
\usepackage{url}
\usepackage{multirow}
\usepackage{graphicx}
\usepackage{epsfig}
\usepackage{epstopdf}
\usepackage{booktabs}
\usepackage{multirow}
\usepackage{siunitx}
\usepackage{adjustbox}
\usepackage{caption}
\usepackage{array}
\usepackage{etoolbox}
\usepackage{graphicx, nicefrac}
\newcolumntype{L}[1]{>{\raggedright\let\newline\\\arraybackslash\hspace{0pt}}m{#1}}
\newcolumntype{C}[1]{>{\centering\let\newline\\\arraybackslash\hspace{0pt}}m{#1}}
\newcolumntype{R}[1]{>{\raggedleft\let\newline\\\arraybackslash\hspace{0pt}}m{#1}}

\begin{document}
\title[$\beta$-decay properties of Y isotopes]
{${\mathbf \beta}$-Decay Properties of Neutron-rich Yttrium Isotopes}

\author{Fakeha Farooq$^1$, Jameel-Un Nabi$^{2,3}$, Ramoona Shehzadi$^1$}
\address{$^1$ Department of Physics, University of the Punjab,
Lahore, Pakistan.}
\address{$^2$ University of Wah, Quaid Avenue, Wah Cantt 47040, Punjab, Pakistan.}
\address{$^3$ Faculty of Engineering Sciences, GIK Institute of Engineering Sciences and Technology, Topi
23640, Khyber Pakhtunkhwa, Pakistan.}

\ead{ramoona.physics@pu.edu.pk}

\begin{abstract}
In this study, we have reported key nuclear properties of weak
$\beta$-decay processes on Yttrium isotopes for the mass number
range A $= 101-108$. This mass region has importance while dealing
with astrophysical r-process abundances. Our study might be helpful
in r-process simulations. We have computed charge-changing strength
distributions, $\beta$-decay half-lives, $\beta$-delayed neutron
emission probabilities and $\beta^{-}$ (EC) weak rates under stellar
conditions. We have performed microscopic calculations based on
deformed proton-neutron quasi-particle
random phase approximation
(pn-QRPA) over a wide temperature ($10^{7} - 3 \times 10^{10}$) K
and density ($10-10^{11}$) g/cm$^{3}$ domain. Unique
first-forbidden (U1F) transitions have been included in the
calculations in addition to the allowed transitions. A significant
decrease in calculated half-lives in ceratin cases, e.g., in $^{107}$Y ($^{108}$Y) by about 67\% (42\%), has been
observed because of the contribution from U1F transitions. We
have compared present results with measured and theoretical works. A good
agreement of our half-lives with experimental
data is observed.
\end{abstract}


\submitto{\PS}
\maketitle

\section{Introduction}
\label{sec:intro}

Since the preliminary work done
by Hoyle~\cite{Hoy46}, Hoyle \etal~\cite{Hoy56}, Cameron~\cite{Cam57} and Burbidge \etal~\cite{Bur57},
nucleosynthesis in high-mass stars has been an active
field of study in nuclear astrophysics (for recent reviews, see e.g.,~\cite{Cyb16,Joh19}).
One of the main unresolved issues in
this field involves the origin of elements heavier than
$^{56}$Fe. Nearly one half of the elemental abundance
beyond iron is created by the s-process (where 's' refers to
slow neutron-capture) of nucleosynthesis~\cite{Kap89,Kap11}, during
the helium burning phases of the stars. For the generation
of the other half of the abundance, r-process (known as rapid neutron-capture) is cited to be the possible
mechanism~\cite{Hoy56,Sat74,Cow91,Amo07,Horo19,Cow21}. Whereas, the
astrophysical location for r-process has not been determined yet, the
process is considered to happen in explosive environments prevailing
the conditions of extreme temperatures ($> 10^{9}$ K) and high
neutron density ($> 10^{20}$ g/cm$^{3}$). Owing to these extreme
physical conditions needed for the r-process, the neutron star mergers~\cite{Wat19} and the core-collapse
supernova are expected to be the most
likely candidates.

During the r-process, the neutron capture occurs at a rate much
faster as compared to the competing $\beta$-decay, leading to the
formation of very high neutron-rich isotopes by a sequence of
neutron captures until there comes the neutron-drip line.
The neutron-rich isotope may then undergo several $\beta$-decays
until a stable isotope is formed on which the capturing process of
neutrons will begin again. The comprehension of the r-process is very important to understand the elemental abundance in the universe.
A proper modeling and simulation of the
r-process is quite challenging and demands an accurate information
of the nuclear characteristics of the neutron-abundant nuclides
across the r-process path. The properties of nuclear
$\beta$-decay~\cite{Kra93}, including half-lives of $\beta$-decay
(T$_{\beta_{1/2}}$), the emission probabilities of beta-delayed neutrons (P$_{n}$), and nuclear masses
are the key ingredients required to describe the r-process nucleosynthesis
and to calculate the final distribution of elemental abundances, which is then compared with
the astronomical observations~\cite{Arc11,Mum16}. The masses of the nuclei are important to define the possible paths of the r-process along the neutron-drip line.
The values
of T$_{\beta_{1/2}}$ of neutron-abundant nuclei across the r-process track determine
the relative abundances and the speed of the matter flow towards the heavier
nuclei, and hence set the time-scale of the process. The values of P$_{n}$ of
the r-process nuclei are particularly important at late stages where
$\beta$-decay competes with the neutron capture, and hence specify the decay path pointing to the stability.

The quality of the nuclear data which is used as an input directly
affects the quality of the nucleosynthesis modeling. Although much
efforts are being done recently, yet most of the nuclear structure
details of neutron-abundant nuclei pertinent for the r-process are
still not experimentally available. The situation is anticipated to
become better with the future experiments involving beams of exotic
nuclei at GSI~\cite{Kurt09} and RIKEN~\cite{Nish11,Hal21}. For now,
the simulation of the r-process primarily relies on theoretically
predicted nuclear data. An extensive study involving the microscopic
computations of the beta-decay half-lives for a very large number of
neutron-abundant nuclei was done by Klapdor-Kleingrothaus
\etal~\cite{Klpa1984}. Later, the proton-neutron quasi particle
random phase approximation (pn-QRPA) model was employed by Staudt
{\it et al}~\cite{Sta89,Sta90} and Hirsch \etal~\cite{Hir93} to
predict T$_{\beta_{1/2}}$ for a vast range of neutron-rich and
proton-rich nuclei. Afterwards, this model has been implemented by
Nabi \etal~\cite{Nabi04} for weak rates calculations of neutron-rich
isotopes of several elements having A$\leq100$. In his
recent publications, Nabi {\it et al} presented $\beta$-decay
half-lives, GT strength distributions, phase space and stellar weak
rates of waiting point nuclei having N = 50,
82~\cite{Nabi19,Nabi2021} and N = 126~\cite{Nabi21} by employing the
deformed pn-QRPA model. The $\beta$-decay properties of even-even
chromium isotopes were earlier studied using the same
model~\cite{Nabi2020}. The model uses multi-shell model space
extended upto 7-$\hbar\omega$ and hence, can accommodate arbitrarily
high-mass nuclei in the calculations. In addition, it calculates
Gamow-Teller (GT) strengths by using state-by-state microscopic
approach for all parent nucleus excited states in contrast to
Independent Particle Model (IPM) ~\cite{Fuller} and large-scale
shell model (LSSM)~\cite{Lang00,LangMar03}.

Presently, we focus on the calculations of weak-decay properties of some
important neutron-abundant nuclei with $100 \le A < 110$. This intermediate
mass region is of particular significance in regards to the astrophysical r-process.
The investigation of their T$_{\beta_{1/2}}$ might be helpful in
the prediction of half-lives of exotic nuclei.
In the studies done by Sarriguren and Pereira~\cite{Sar10} and Sarriguren \etal~\cite{Sar14}, the $\beta$-decay properties of several
Zirconium and Molybdenum isotopes in the mass region 100-110 have been reported. In a work, related to universality of r-process~\cite{Lor15}, the $\beta$-decay half-lives of $^{104-108}$Y
isotopes were measured at the Radioactive Isotope Beam Factory. Theoretical work based on an empirical formula for the calculations T$_{\beta_{1/2}}$ of
$^{101-109}$Y has recently been performed by~\cite{Zhi18}. We have also chosen Y isotopes, $^{101-108}$Y, for the present analysis.
We have computed terrestrial half-lives of $\beta$-transitions, $\beta$-delayed
neutron emission probabilities, charge-changing strength-distributions
and stellar weak ($\beta^{-}$ and EC) rates by employing deformed pn-QRPA model.
In addition, we have compared our calculated results with the experimental data and with
earlier theoretical estimations (for detail, see section~\ref{sec:results}).
The present calculations include contributions from both
unique-first forbidden (U1F) transitions and allowed GT. As per previous studies,
the inclusion of first-forbidden (FF) transitions to the calculations of
$\beta$-decay properties greatly impacts the half-lives of nuclei, especially,
near-stable, and near-magic number nuclei~\cite{Hom96} and those of
r-process nuclei~\cite{ Mol03}. For the inclusion of non-unique FF transitions
in the calculations, work on code is in process.

The current paper
is organised as follows: in section~\ref{sec:formalism} the
formalism based on pn-QRPA theory has been discussed. Sections~\ref{sec:results} and
\ref{sec:conclusions} present results and discussion of our study, and conclusions based on our findings, respectively.

\section{Formalism}
\label{sec:formalism}
The study of nuclear properties associated with weak $\beta$-decay processes,
especially of those nuclei which lie far beyond the valley of stability,
has considerable importance in understanding the phenomena happening
in astrophysical environments. There is a need of reliable theoretical estimation of these
properties of such heavier nuclei. The pn-QRPA is a simple but microscopic theoretical
model having wide applicability in astrophysical calculations.
In this model, the eigenvectors and energies of single-particle states of quasi-particle system are estimated by the deformed Nilsson model~\cite{Nil55}, with a well-established oscillatory potential (OP) having
quadratic deformation. Other required parameters adopted from the Nilsson model are;
the Nilsson potential parameters, and the Nilsson oscillator constant, which is
taken as $\hbar\omega_{o} = \nicefrac{41}{A^{1/3}}\;$(MeV). The BCS approximation is
applied for the treatment of pairing correlations between nucleons with constant pairing
forces. The deformed Nilsson basis are used to accomplish
BCS calculations for the systems of protons and neutrons, separately. A residual interaction is included
for the correlations in ground-state through
particle-to-hole (ph) and particle-to-particle (pp) GT forces, which are treated in RPA. The pp and ph GT force strength parameters are denoted by $\kappa_{GT}$ and
$\chi_{GT}$, respectively.
The details of the formalism used for the calculation of allowed
$\beta$-decay rates in stellar environments can be seen from~\cite{Nabi04,Nabi99}.

In the case of U1F transitions, the ph and pp matrix
elements are
\begin{equation}
V^{ph}_{pn,p^{\prime}n^{\prime}} = +2\chi_{U1F}f_{pn}(\mu)f_{p^{\prime}n^{\prime}}(\mu)
\label{phF}
\end{equation}
\begin{equation}
V^{pp}_{pn,p^{\prime}n^{\prime}} = -2\kappa_{U1F}f_{pn}(\mu)f_{p^{\prime}n^{\prime}}(\mu)
\label{ppF}
\end{equation}
where
\begin{equation}
f_{pn}(\mu)=\langle p|t_{-}r[\sigma_{\mu}Y_{1}]_{2\mu}|n \rangle
\end{equation}
is a single-particle U1F transition amplitude. $\mu$ denotes the spherical component of the transition operator and has values $0,\pm1,\pm2$. Other symbols have their usual meanings. The proton ($ |p \rangle$)
and neutron ($|n \rangle$) states have opposite parities~\cite{Hom96}.

The strength parameters are tuned
so that the computations reproduce the experimentally observed~\cite{Aud17}
$\beta$-decay half-lives. We employed
$\kappa_{GT} = 0.25/A^{0.7}\;$(MeV) and $\chi_{GT} = 2.8/A^{0.7}\;$(MeV) in our calculations
for allowed and $\kappa_{U1F} = 0\;$(MeV) and $\chi_{U1F} = 71/A\;$(MeV)
for forbidden transitions~\cite{Hom96}.
The choice of deformation parameter ($\varepsilon_{2}$) is
made such that it has quadrupole dependence~\cite{Mol81}.
\begin{equation}
\varepsilon_{2} = RQA^{\nicefrac{-2}{3}}Z^{-1};~~~~R=\frac{125}{1.44},
\label{Eq:Dpar}
\end{equation}
where $Q$, $A$ and $Z$ are the electric
quadrupole moment, atomic mass and atomic number, respectively. The BCS approximation
calculations in the present work is based on the constant pairing
forces between the proton (neutron) system having fixed pairing
strength with pairing energy gaps $\Delta p$ ($
\Delta n$). The following globally recognized
expressions~\cite{Boh69} for these pairing gaps are used,
\begin{equation}
\Delta n = \Delta p = \frac{12}{\sqrt{A}}~~~~(MeV).
\label{Eq:pairgap}
\end{equation}

Experimental mass excess values reported in Audi \etal~\cite{Aud17} are used to calculate Q-values.
For a comprehensive
description of this extended model including separable GT-interactions, the reader is
referred to~\cite{Muto92}.
The half-life of a $\beta$-transition can be obtained by,
\begin{equation}
T_{\beta_{\nicefrac{1}{2}}}=\left(\sum_{0 \leq E_{j} \leq Q_{\beta}}\frac{1}{t_{j}}\right)^{-1},
\label{Eq:HL}
\end{equation}
where sum is taken over all partial probabilities of $\beta$-transitions from initial level $i^{th}$
of parent nucleus to the daughter $j^{th}$ level having energies ($E_{j}$) within
the $Q_{\beta}$-windows. The partial half-life ($t_{j}$) is given by,
\begin{equation}
t_{j}=\frac{D}{F_{V}(Z;Q_{\beta}-E_{j})B_{F}(E_{j})+(\nicefrac{g_{A}}{g_{V}})^{2}F_{A}(Z;Q_{\beta}-E_{j})B_{GT}(E_{j})}.
\label{Eq:PHL}
\end{equation}

The ratio between coupling constants (axial-vector, $g_{A}$, to vector, $g_{V}$)~\cite{Nak10} and constant D~\cite{Har09} take values,
\begin{equation}
\frac{g_{A}}{g_{V}}=-1.2694;~~~~D=6143~s.
\end{equation}

In~\eref{Eq:PHL}, $F_{V}$ and $F_{A}$ are integrated phase space Fermi functions~\cite{Gove71}.
$B_{F}(E_{j})$ and $B_{GT}(E_{j})$ are the reduced probabilities of transitions associated with
Fermi and GT-decays, respectively, in $\beta^{\pm}$ directions and are given by,
\begin{eqnarray}
B(F_{\pm};ij) &=& \frac{|\langle j||\sum_{l}t^{l}_{\pm}||i \rangle|^{2}}{2J_{i}+1},
\label{Eq:FTP}
\end{eqnarray}
\begin{eqnarray}
B(GT_{\pm};ij) &=& \frac{|\langle j||\sum_{l}t^{l}_{\pm}\vec{\sigma}^{l}||i \rangle|^{2}}{2J_{i}+1},
\label{Eq:GTP}
\end{eqnarray}
where $J_{i}$, $t^{l}_{\pm}$ and $\vec{\sigma}^{l}$ are the
total $i^{th}$ state spin of nucleus, raising (lowering) isospin operators and operator
associated with Pauli spin matrices, respectively.
The spin and isospin operators act on the $l^{th}$ nucleon of a nucleus.

Decay rates or probabilities of stellar $\beta$-transitions, from
initial level ($i^{th}$) of parent nucleus to final level ($j^{th}$)
of daughter nucleus, are calculated by,
\begin{equation}
\lambda_{ij} = \left(\frac{m^{5}_{e}c^{4}}{2\hbar^{7} \pi^{3}} \right)
\sum_{\Delta J^{\pi}}g^{2}f(\Delta J^{\pi};ij)
B(\Delta J^{\pi};ij)
\label{Eq:DR}
\end{equation}

This equation contains the Fermi integrals ($f$) and reduced transition probability
($B$) of a decay which give rise to a spin-parity ($\Delta J^{\pi}$) change.
The weak coupling constant ($g$) takes the value $g_{A}$ ($g_{V}$)
for transition governed by axial-vector (vector) weak-force. This equation deals with
both allowed GT-transitions and forbidden decays (FD). In case of allowed transitions,
contributions only come from vector type Fermi (F) transitions ($\Delta J^{\pi}=0^{+}$)
and axial-vector type GT transitions ($\Delta J^{\pi}=1^{+}$). The reduced probabilities of Fermi and GT-transitions are given by~\eref{Eq:FTP} and
\eref{Eq:GTP}, respectively. The Fermi integrals $f(\Delta J^{\pi};ji)$
in case of $\beta^{-}$-decay and for continuum capturing of electrons are given by,
\begin{eqnarray}
f^{\beta^{-}}(\Delta J^{\pi};ij) &=& \int_{1}^{\epsilon_{m}}\epsilon({\epsilon^{2} -1)^{1/2}(\epsilon_{m}-\epsilon)^{2}
F(+Z,\epsilon)(1-G_{-})}d\epsilon,
\label{Eq:FIEE}
\end{eqnarray}
\begin{eqnarray}
f^{EC}(\Delta J^{\pi}};ij) = \int _{\epsilon_{l}}^{\infty}{\epsilon
(\epsilon^{2} -1)^{1/2}(\epsilon_{m}+\epsilon)^{2}
F(+Z,\epsilon)G_{-}\ d\epsilon.
\label{Eq:FIPC}
\end{eqnarray}

In~\eref{Eq:FIEE} and \eref{Eq:FIPC}, $\epsilon_{m}$, $\epsilon_{l}$ and $\epsilon$ are the total $\beta$-transition energy, electron capture threshold energy and total energy of electron (sum of rest mass and  kinetic energy),
respectively. The $G_{-}$ are the Fermi Dirac distribution functions for electron.

For the U1F decays, reduced probability and Fermi integrals are calculated according
to the following equations,
\begin{equation}
B(U1F;ij) = \frac{1}{12}x^{2}(\epsilon^{2}_{m}-1)
-\frac{1}{6}x^{2}\epsilon_{m}\epsilon+
\frac{1}{6}x^{2}\epsilon^{2},
\label{Eq:UFTP}
\end{equation}

where $x$ is given by,
\begin{eqnarray}
x &=& 2g_{A}\frac{1}{\sqrt{2J_{i}+1}}\langle j||\sum_{l}r_{l}[\mathbf{C}^{l}_{1}\times\vec{\sigma}]^{2}t^{l}_{-}||i \rangle, \\
\mathbf{C}_{mm^{'}} &=& \left(\frac{4\pi}{2l+1}\right)^{1/2}\mathbf{Y}_{mm^{'}},
\label{Eq:z}
\end{eqnarray}

where $\mathbf{Y}_{\mathtt{mm^{'}}}$ are the spherical harmonics.
\begin{eqnarray}
f(\Delta J^{\pi};ij) &=& \int _{1}^{\epsilon_{m}}\{\epsilon
(\epsilon^{2} -1)^{1/2}(\epsilon_{m}-\epsilon)^{2} \\ \nonumber
&~&[(\epsilon_{m}-\epsilon)^{2} F_{1}(Z,\epsilon)+(\epsilon^{2}-1)F_{2}(Z,\epsilon)](1-G_{-})\}d\epsilon
\label{Eq:UFI}
\end{eqnarray}
Fermi functions; F$_{1}$ and F$_{2}$, are adopted from Gove at al.~\cite{Gove71}, like in the case of allowed transitions.
The total decay rates per nucleus are calculated as,
\begin{equation}
\lambda_{\beta^{-}(EC)}\equiv\lambda(\beta) = \sum _{ij}P_{i} \lambda(\beta;ij),
\label{Eq:Trate}
\end{equation}
here $P_{i}$ is the occupation probability of  excited states of
parent nucleus obeying the normal Boltzmann distribution. The
summation in~\eref{Eq:Trate} is performed over all initial (parent)
states as well as over all final (daughter) states to obtain desired
convergence in the calculated rate values. In our
calculations, it was further assumed that if the excited states
energies ($E_j$) of daughter nuclei are greater than the neutron
separation energy ($S_n$), nuclei decay through neutron emission. In
this case, neutron emission energy rate are computed by,
\begin{equation}
\lambda_{n}=\sum_{ij}P_i\lambda_{ij}(E_{j}-S_n);    \qquad \forall E_{j}>S_{n} .
\label{Eq:Pn1}
\end{equation}
The $\beta$-delayed neutron emission probability ($P_{n}$) is calculated by,
\begin{equation}
P_n=\frac{\sum_{ij^\prime}{P_i\lambda_{ij^\prime}}}{\sum_{ij}{P_i\lambda_{ij}}} ,
\label{Eq:Pn2}
\end{equation}
In ~\eref{Eq:Pn1} and ~\eref{Eq:Pn2}, $\lambda_{n}$ is the total
energy rates for neutron emission and $\lambda_{ij}$ is the partial
rate for an $i$ state to $j$ state transition, which is the
combination of positron capture and $\beta^{-}$-decay rates. The
index $j^{\prime}$ represents the energy levels having $E_{j} >
S_{n}$ of the daughter nucleus.

\section{Results and discussions}
\label{sec:results}

The pn-QRPA calculated terrestrial half-lives associated with weak $\beta$-decay
(T$_{\beta_{1/2}}$) and $\beta$-delayed
neutron emission probabilities (P$_{n}$) for $^{101-108}$Y have been reported in this
section. We have also presented a comparison of our findings with experimentally measured
half-lives~\cite{Aud17} and with pioneering calculations of M\"{o}ller \etal
(using QRPA + gross-theory)~\cite{Mol03} and Pfeiffer \etal (using QRPA-1, QRPA-2 and KHF)~\cite{Pfe02}. Moreover, we have displayed
our calculated charge-changing strength distributions and stellar weak rates for both
U1F and allowed GT transitions, separately in $\beta^{-}$ and EC directions. In the
calculations of weak rates, stellar temperature range from $10^{7}$ K to
$3 \times 10^{10}$ K and density range from $10$ g/cm$^{3}$ to $10^{11}$ g/cm$^{3}$ have
been considered.

\begin{table}
\caption{\small Comparison of pn-QRPA (This work) calculated $\beta$-decay
half-lives (in units of second) with theoretical results of
~\cite{Mol03} (QRPA + gross-theory), and~\cite{Pfe02}
(QRPA-1, QRPA-2 and KHF) and with
experimental data \cite{Aud17}.}\label{table1}
\centering {\resizebox{!}{3cm}{%
\begin{tabular}{lccccccc}
\toprule
\multirow{2}{*}{Y isotopes} &
\multicolumn{7}{c}{T$_{\beta_{1/2}} [s]$} \\
\cmidrule(l){2-3}  \cmidrule(l){3-5}  \cmidrule(l){5-8}
& Experimental & \multicolumn{2}{c}{pn-QRPA} & QRPA+gross theory & KHF & QRPA-1 & QRPA-2 \\
&          & (GT) &  (GT+FF) & (GT+FF) &       &       &  \\
\cmidrule(l){1-3}  \cmidrule(l){3-5}  \cmidrule(l){5-8}
    $^{101}$Y  & 0.426 & 0.533 & 0.481 & 0.194 & 0.325 & 0.149 & 0.149 \\
    $^{102}$Y  & 0.298 & 0.314 & 0.306 & 0.107 & 0.352 & 0.189 & 0.189 \\
    $^{103}$Y  & 0.239 & 0.300 & 0.209 & 0.087 & 0.181 & 0.089 & 0.089 \\
    $^{104}$Y  & 0.197 & 0.201 & 0.194 & 0.032 & 0.127 & 0.030  & 0.030 \\
    $^{105}$Y  & 0.095 & 0.136 & 0.083 & 0.048 & 0.088 & 0.048 & 0.048 \\
    $^{106}$Y  & 0.074 & 0.103 & 0.081 &  ---     & 0.066 & 0.035 & 0.035 \\
    $^{107}$Y  & 0.034 & 0.080 & 0.026 &  ---   & 0.074 & 0.031 & 0.031 \\
    $^{108}$Y  & 0.030  & 0.041 & 0.023 & ---    & 0.048 & 0.023 & 0.023 \\
\bottomrule
\end{tabular}}
}\end{table} In table~\ref{table1}, the column labelled as, pn-QRPA,
presents the terrestrial T$_{\beta_{1/2}} [s]$ values calculated in
this work by taking contributions from both U1F and allowed GT
transitions. It can be seen from the table that our results show
reasonable agreement with experimental half-life values~\cite{Aud17}
after the inclusion of U1F. Table~\ref{table1} also shows the
comparison of our results with previous calculations. In QRPA+gross
theory calculations~\cite{Mol03}, the allowed GT calculations were
performed within the QRPA framework and the inclusion of FF
transitions was done by using gross theory. Pfeiffer
\etal~\cite{Pfe02} have reported the results of unified
macroscopic-microscopic model within the quasi-particle random-phase
approximation~\cite{Krum84,Mol90} (named as QRPA-1 and QRPA-2) and
using an updated form of the empirical Kratz Herrmann formula (KHF).
In both QRPA-1 and QRPA-2, authors used the latest mass evaluation
values available at that time and they transformed all the
GT-transition strengths above 2 MeV to a Gaussian of width $\Delta =
8.62/A^{0.57}$. However, in QRPA-2, additionally, the ground state
deformations were considered in a more accurate way. A nucleus known
to be spherical in accordance with experiment was treated as
spherical, although, a deformed shape was obtained in the FRDM mass
model~\cite{Mol92} calculations of ground state deformations. KHF is
a simple phenomenological expression for delayed-neutron emission
probability, $P_{n}$, which was initially developed by Kratz and
Herrman in 1972~\cite{Kratz73}. Later, Pfeiffer and
collaborators~\cite{Pfe00} suggested to fit the T$_{1/2}$ of
neutron-rich nuclides according to the underlying concepts of KHF
and developed an expression for half-life calculations. Based on new
non-linear least-squares fit parameters, with the inclusion of new
experimental values available at that time, the authors Pfeiffer
\etal~\cite{Pfe02} used KHF to predict the unknown $P_{n}$ values
and unknown T$_{1/2}$. The authors only computed the allowed GT
rates in their QRPA calculations. Table~\ref{table2} depicts the
comparison of our calculated P$_{n}$ with earlier results. Our
calculated probabilities are notably different from the other model
computations. The primary reason for this difference might
be attributed to the different way of the calculations of daughter
energy levels and matrix elements of charge-changing transitions
between the initial and final states by the different models.
\begin{table}
\caption{\small Comparison of pn-QRPA (This work) calculated P$_{n}$ with theoretical results of
~\cite{Mol03} (QRPA + gross-theory) and~\cite{Pfe02}
(QRPA-1, QRPA-2 and KHF).}\label{table2}
\centering {\resizebox{!}{3cm}{%
\begin{tabular}{lccccc}
\toprule
\multirow{2}{*}{Y isotopes} &
\multicolumn{5}{c}{P$_{n} (\%)$} \\
\cmidrule(l){2-3}  \cmidrule(l){3-6}
& pn-QRPA & QRPA+gross theory & KHF & QRPA-1 & QRPA-2 \\
\cmidrule(l){1-3}  \cmidrule(l){3-6}
   $^{101}$Y  & 1.0 & 0.6 & 3.9 & 1.2 & 1.2 \\
   $^{102}$Y & 10.5 & 3.4 & 3.7 & 1.2 & 1.2 \\
   $^{103}$Y  & 23.7 & 4.0 & 8.5 & 3.5 & 3.5 \\
   $^{104}$Y  & 44.1 & 4.8 & 11.6 & 3.2 & 3.2 \\
   $^{105}$Y  & 32.5 & 22.4 & 20.4 & 14.0 & 14.0 \\
   $^{106}$Y  & 65.7 & ---  & 24.0 & 16.3 & 16.3 \\
    $^{107}$Y  & 35.0 & ---  & 31.7 & 32.1 & 32.1 \\
   $^{108}$Y  &65.2  &  --- & 25.5 & 36.0 & 36.0 \\
\bottomrule
\end{tabular}}
}\end{table}

\begin{figure}[h]
\begin{center}
\includegraphics[width=1.2\textwidth]{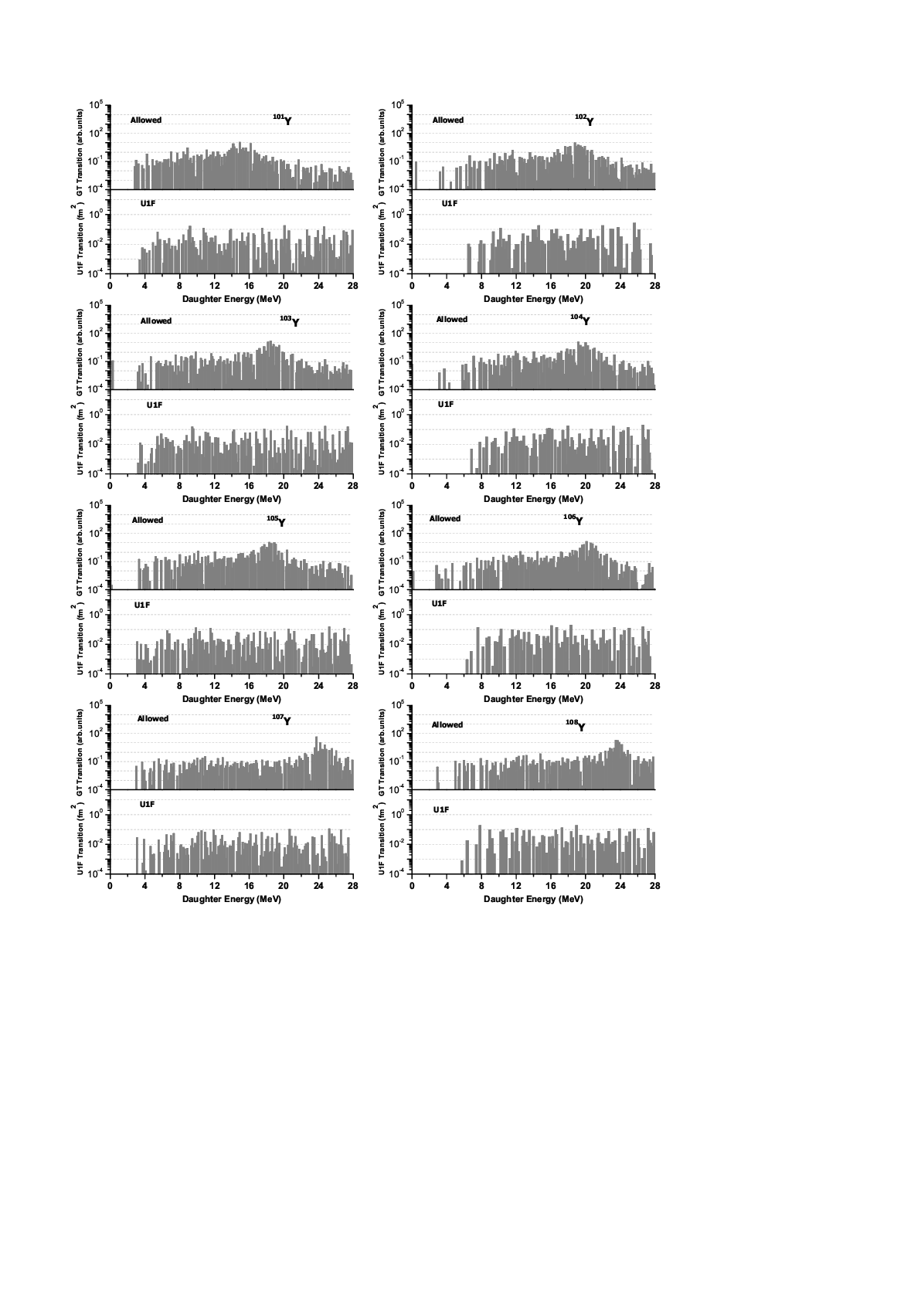}
\vspace{-9cm}\caption{Charge-changing strength distributions of $^{101-108}$Y for allowed and
unique-first forbidden (U1F)
transitions in $\beta^{-}$ direction .}
\label{figure1}
\end{center}
\end{figure}
The charge-changing strength distributions for allowed and U1F
transitions have been computed with respect to parent ground, and
excited states. However, due to space consideration, only the
results of computed strengths in case of transitions from parent
ground level to daughter ground and excited levels, up to 28 MeV
excitation energy of daughter states, have been reported. The ASCII
files of these strength distributions are available and may be
requested from the corresponding author. The graphs of
charge-changing strength distributions of $^{101-108}$Y, in
$\beta^{-}$ direction, have been shown in figure~\ref{figure1} for
allowed, as well as U1F decays. The upper and bottom panels in case
of each isotope show the allowed GT and U1F contributions,
respectively. Relatively large contributions of U1F transitions can
be noted in case of $^{107, 108, 105, 103}$Y as compared to $^{101,
102, 104, 106}$Y. The large U1F contribution for A = 103,
105, 107 and 108 isotopes is attributed to the fact that the
available phase space of U1F transition is bigger than allowed GT
phase space in all these cases in comparison to
$^{101,102,104,106}$Y nuclei. Therefore, the incorporation of U1F
transitions in our calculations reduces the terrestrial half-lives
of $^{107}$Y, $^{108}$Y, $^{105}$Y and $^{103}$Y by about 67$\%$,
42$\%$, 38$\%$ and 30$\%$, respectively. However, for $^{101}$Y
($\sim$ 9.7$\%$), $^{102}$Y ($\sim$ 2.6$\%$), $^{104}$Y ($\sim$
3.75$\%$) and $^{106}$Y ($\sim$ 21.32$\%$), relatively small
reduction in T$_{\beta_{1/2}}$ is observed.

\begin{table}[tp]
\caption{\small The pn-QRPA calculated $\beta^{-}$ rates ($\lambda_{\beta^{-}}$)
due to allowed transitions on $^{101-108}$Y at various selected
densities and temperatures in stellar environment. The calculated
rates are in logarithmic (to the base 10) scale in units of s$^{-1}$.
$\rho \text{Y}_{e}$ has units of gcm$^{-3}$, where $\rho$ is the baryon
density and Y$_{e}$ is the ratio of the lepton number to the baryon
number. Temperature (T$_{9}$) is given in units of $10^{9}$\;K. Here -100 indicates that the computed rate is less than 1.00$\times$10$^{-100}$ s$^{-1}$.}\label{table3} \centering {\resizebox{!}{10.7cm}{%
\begin{tabular}{cccccccccccc}
\toprule \multirow{2}{*}{$\rho$Y$_{e}$} &
\multirow{2}{*}{T$_{9}$} & \multicolumn{8}{c}{$\lambda_{\beta^{-}} [s^{-1}]$ (allowed)} \\
\cmidrule{3-4}  \cmidrule{5-6}  \cmidrule{7-8} \cmidrule{9-10}& &
\multicolumn{1}{c}{$^{101}$Y} &
\multicolumn{1}{c}{$^{102}$Y} &
\multicolumn{1}{c}{$^{103}$Y} &
\multicolumn{1}{c}{$^{104}$Y} &
\multicolumn{1}{c}{$^{105}$Y} &
\multicolumn{1}{c}{$^{106}$Y} &
\multicolumn{1}{c}{$^{107}$Y} &
\multicolumn{1}{c}{$^{108}$Y} \\
\midrule
10$^{3}$ & 0.4   & 0.144 & 0.345 & 0.461 & 0.551 & 0.715 & 1.165 & 0.957 & 1.450 \tabularnewline
10$^{3}$& 0.7   & 0.174 & 0.407 & 0.479 & 0.599 & 0.728 & 1.249 & 1.017 & 1.522 \tabularnewline
10$^{3}$& 1     & 0.203 & 0.450  & 0.502 & 0.661 & 0.747 & 1.297 & 1.066 & 1.548 \tabularnewline
10$^{3}$& 1.5   & 0.250  & 0.507 & 0.539 & 0.752 & 0.78  & 1.342 & 1.115 & 1.561 \tabularnewline
10$^{3}$ & 2     & 0.290  & 0.554 & 0.568 & 0.815 & 0.807 & 1.361 & 1.145 & 1.563 \tabularnewline
10$^{3}$  & 3     & 0.346 & 0.630  & 0.602 & 0.888 & 0.843 & 1.372 & 1.178 & 1.562 \tabularnewline
10$^{3}$  & 5     & 0.425 & 0.730  & 0.648 & 0.959 & 0.898 & 1.378 & 1.227 & 1.580 \tabularnewline
10$^{3}$  & 10    & 0.966 & 1.006 & 1.099 & 1.227 & 1.342 & 1.575 & 1.661 & 1.878 \tabularnewline
10$^{3}$  & 15    & 1.527 & 1.364 & 1.606 & 1.592 & 1.821 & 1.896 & 2.104 & 2.285 \tabularnewline
10$^{3}$  & 20    & 1.850  & 1.598 & 1.908 & 1.833 & 2.104 & 2.111 & 2.360  & 2.535 \tabularnewline
10$^{3}$  & 25    & 2.037 & 1.736 & 2.085 & 1.979 & 2.271 & 2.240  & 2.509 & 2.68 \tabularnewline
10$^{3}$  & 30    & 2.153 & 1.820  & 2.195 & 2.070  & 2.374 & 2.320  & 2.601 & 2.769 \tabularnewline
          &       &       &       &       &       &       &       &       & \tabularnewline
10$^{6}$  & 0.4   & 0.136 & 0.339 & 0.456 & 0.547 & 0.710  & 1.162 & 0.954 & 1.448 \tabularnewline
10$^{6}$   & 0.7   & 0.167 & 0.401 & 0.475 & 0.595 & 0.724 & 1.246 & 1.014 & 1.520\tabularnewline
10$^{6}$  & 1     & 0.197 & 0.445 & 0.498 & 0.658 & 0.744 & 1.294 & 1.063 & 1.546 \tabularnewline
10$^{6}$  & 1.5   & 0.245 & 0.503 & 0.536 & 0.749 & 0.777 & 1.339 & 1.114 & 1.559 \tabularnewline
10$^{6}$  & 2     & 0.286 & 0.551 & 0.565 & 0.813 & 0.804 & 1.359 & 1.143 & 1.561 \tabularnewline
10$^{6}$  & 3     & 0.343 & 0.628 & 0.600   & 0.887 & 0.842 & 1.371 & 1.177 & 1.561 \tabularnewline
10$^{6}$  & 5     & 0.424 & 0.729 & 0.647 & 0.958 & 0.897 & 1.377 & 1.226 & 1.579 \tabularnewline
10$^{6}$  & 10    & 0.965 & 1.005 & 1.099 & 1.226 & 1.342 & 1.574 & 1.661 & 1.878 \tabularnewline
10$^{6}$  & 15    & 1.527 & 1.364 & 1.606 & 1.592 & 1.821 & 1.896 & 2.103 & 2.285 \tabularnewline
10$^{6}$  & 20    & 1.850  & 1.598 & 1.908 & 1.833 & 2.104 & 2.111 & 2.360  & 2.535 \tabularnewline
10$^{6}$  & 25    & 2.037 & 1.736 & 2.085 & 1.979 & 2.271 & 2.240  & 2.509 & 2.680 \tabularnewline
10$^{6}$  & 30    & 2.153 & 1.820  & 2.195 & 2.070  & 2.374 & 2.320  & 2.601 & 2.769 \tabularnewline
          &       &       &       &       &       &       &       &       &  \tabularnewline
10$^{9}$  & 0.4   & -1.29 & -0.644 & -0.589 & -0.1  & -0.2  & 0.552 & 0.296 & 0.928 \tabularnewline
10$^{9}$  & 0.7   & -1.004 & -0.565 & -0.428 & -0.072 & -0.093 & 0.636 & 0.419 & 1.006 \tabularnewline
10$^{9}$  & 1     & -0.871 & -0.504 & -0.334 & -0.031 & -0.019 & 0.688 & 0.509 & 1.043 \tabularnewline
10$^{9}$  & 1.5   & -0.735 & -0.419 & -0.231 & 0.052 & 0.067 & 0.736 & 0.593 & 1.070 \tabularnewline
10$^{9}$  & 2     & -0.652 & -0.346 & -0.17 & 0.121 & 0.121 & 0.758 & 0.637 & 1.079 \tabularnewline
10$^{9}$  & 3     & -0.551 & -0.23 & -0.103 & 0.211 & 0.186 & 0.774 & 0.685 & 1.089\tabularnewline
10$^{9}$  & 5     & -0.394 & -0.049 & -0.009 & 0.327 & 0.286 & 0.808 & 0.762 & 1.13 \tabularnewline
10$^{9}$  & 10    & 0.469 & 0.470  & 0.641 & 0.765 & 0.914 & 1.147 & 1.323 & 1.552 \tabularnewline
10$^{9}$  & 15    & 1.221 & 1.019 & 1.306 & 1.284 & 1.536 & 1.607 & 1.869 & 2.066 \tabularnewline
10$^{9}$  & 20    & 1.658 & 1.381 & 1.716 & 1.636 & 1.919 & 1.924 & 2.203 & 2.390 \tabularnewline
10$^{9}$  & 25    & 1.917 & 1.599 & 1.963 & 1.853 & 2.152 & 2.121 & 2.407 & 2.585 \tabularnewline
10$^{9}$  & 30    & 2.075 & 1.732 & 2.116 & 1.988 & 2.297 & 2.242 & 2.533 & 2.705 \tabularnewline
          &       &       &       &       &       &       &       &       &  \tabularnewline
10$^{11}$  & 0.4   & -100  & -100  & -100  & -100  & -100  & -100  & -100  & -100 \tabularnewline
10$^{11}$ & 0.7   & -100  & -100  & -100  & -100  & -100  & -100  & -86.470 & -88.193\tabularnewline
10$^{11}$ & 1     & -81.140 & -81.606 & -74.801 & -75.807 & -71.006 & -70.953 & -61.410 & -62.546\tabularnewline
10$^{11}$ & 1.5   & -54.934 & -55.208 & -50.679 & -51.268 & -48.176 & -48.040 & -41.773 & -42.454 \tabularnewline
10$^{11}$ & 2     & -41.739 & -41.893 & -38.520 & -38.884 & -36.637 & -36.456 & -31.854 & -32.306 \tabularnewline
10$^{11}$ & 3     & -28.416 & -28.401 & -26.233 & -26.355 & -24.945 & -24.699 & -21.789 & -21.999 \tabularnewline
10$^{11}$ & 5     & -17.446 & -17.309 & -16.197 & -16.104 & -15.364 & -15.048 & -13.490 & -13.453 \tabularnewline
10$^{11}$ & 10    & -8.022 & -8.386 & -7.855 & -7.822 & -7.386 & -7.236 & -6.459 & -6.192 \tabularnewline
10$^{11}$ & 15    & -4.493 & -4.976 & -4.481 & -4.573 & -4.155 & -4.145 & -3.546 & -3.278 \tabularnewline
10$^{11}$ & 20    & -2.657 & -3.144 & -2.664 & -2.798 & -2.403 & -2.442 & -1.940 & -1.692 \tabularnewline
 10$^{11}$ & 25    & -1.531 & -2.010 & -1.537 & -1.689 & -1.309 & -1.373 & -0.924 & -0.695 \tabularnewline
10$^{11}$ & 30    & -0.768 & -1.240 & -0.769 & -0.931 & -0.560 & -0.639 & -0.224 & -0.008 \tabularnewline
 \bottomrule
\end{tabular}}}
\end{table}

\begin{table}[tp]
\caption{\small The same as table \ref{table3} but for EC rates ($\lambda_{EC}$)
due to allowed transitions.}\label{table4} \centering {\resizebox{!}{11.5cm}{%
\begin{tabular}{cccccccccccc}
\toprule \multirow{2}{*}{$\rho$Y$_{e}$} &
\multirow{2}{*}{T$_{9}$} & \multicolumn{8}{c}{$\lambda_{EC} [s^{-1}]$ (allowed)} \\
\cmidrule{3-4}  \cmidrule{5-6}  \cmidrule{7-8} \cmidrule{9-10}& &
\multicolumn{1}{c}{$^{101}$Y} &
\multicolumn{1}{c}{$^{102}$Y} &
\multicolumn{1}{c}{$^{103}$Y} &
\multicolumn{1}{c}{$^{104}$Y} &
\multicolumn{1}{c}{$^{105}$Y} &
\multicolumn{1}{c}{$^{106}$Y} &
\multicolumn{1}{c}{$^{107}$Y} &
\multicolumn{1}{c}{$^{108}$Y} \\
\midrule
10$^{3}$    & 0.4   & -100  & -100  & -100  & -100  & -100  & -100  & -100  & -100 \\
10$^{3}$   & 0.7   & -76.113 & -76.003 & -85.825 & -76.001 & -94.893 & -85.138 & -100  & -87.947 \\
10$^{3}$  & 1     & -54.353 & -54.024 & -61.251 & -54.033 & -67.077 & -60.305 & -70.561 & -62.261 \\
10$^{3}$  & 1.5   & -36.414 & -36.039 & -41.118 & -36.046 & -44.511 & -40.092 & -46.653 & -41.385 \\
10$^{3}$  & 2     & -27.209 & -26.829 & -30.813 & -26.839 & -32.982 & -29.776 & -34.465 & -30.723 \\
10$^{3}$   & 3     & -17.818 & -17.433 & -20.282 & -17.461 & -21.232 & -19.293 & -22.061 & -19.867 \\
10$^{3}$ & 5     & -10.022 & -9.633 & -11.333 & -9.69 & -11.486 & -10.61 & -11.803 & -10.840 \\
10$^{3}$  & 10    & -3.473 & -3.208 & -3.667 & -3.256 & -3.575 & -3.391 & -3.564 & -3.340 \\
 10$^{3}$  & 15    & -0.692 & -0.581 & -0.698 & -0.606 & -0.604 & -0.567 & -0.531 & -0.454 \\
  10$^{3}$  & 20    & 0.916 & 0.947 & 0.944 & 0.934 & 1.028 & 1.003 & 1.112 & 1.133 \\
10$^{3}$  & 25    & 1.973 & 1.960  & 2.011 & 1.955 & 2.086 & 2.025 & 2.171 & 2.162 \\
  10$^{3}$   & 30    & 2.731 & 2.690  & 2.774 & 2.692 & 2.841 & 2.759 & 2.925 & 2.898 \\
          &       &       &       &       &       &       &       &       &  \\
10$^{6}$   & 0.4   & -100  & -100  & -100  & -100  & -100  & -100  & -100  & -100 \\
 10$^{6}$   & 0.7   & -72.490 & -72.381 & -82.203 & -72.378 & -91.270 & -81.515 & -96.515 & -84.325 \\
10$^{6}$ & 1     & -51.196 & -50.867 & -58.094 & -50.876 & -63.920 & -57.148 & -67.404 & -59.104 \\
10$^{6}$ & 1.5   & -34.401 & -34.027 & -39.105 & -34.033 & -42.498 & -38.079 & -44.640 & -39.372 \\
 10$^{6}$ & 2     & -25.921 & -25.541 & -29.525 & -25.551 & -31.695 & -28.488 & -33.177 & -29.436 \\
 10$^{6}$ & 3     & -17.317 & -16.931 & -19.781 & -16.959 & -20.730 & -18.791 & -21.559 & -19.366 \\
10$^{6}$  & 5     & -9.930 & -9.541 & -11.241 & -9.598 & -11.394 & -10.518 & -11.711 & -10.748 \\
 10$^{6}$  & 10    & -3.463 & -3.198 & -3.657 & -3.246 & -3.565 & -3.381 & -3.554 & -3.330 \\
 10$^{6}$   & 15    & -0.689 & -0.578 & -0.696 & -0.603 & -0.601 & -0.564 & -0.528 & -0.451 \\
10$^{6}$   & 20    & 0.917 & 0.948 & 0.946 & 0.935 & 1.029 & 1.004 & 1.114 & 1.134 \\
 10$^{6}$  & 25    & 1.974 & 1.960  & 2.012 & 1.955 & 2.086 & 2.026 & 2.172 & 2.163 \\
 10$^{6}$  & 30    & 2.732 & 2.691 & 2.774 & 2.692 & 2.842 & 2.759 & 2.925 & 2.899 \\
          &       &       &       &       &       &       &       &       &  \\
10$^{9}$   & 0.4   & -68.544 & -69.612 & -85.284 & -69.51 & -100  & -85.800 & -100  & -90.813 \\
10$^{9}$   & 0.7   & -40.281 & -40.139 & -49.985 & -40.136 & -59.028 & -49.273 & -64.273 & -52.083 \\
10$^{9}$  & 1     & -28.511 & -28.169 & -35.408 & -28.178 & -41.221 & -34.449 & -44.705 & -36.406 \\
10$^{9}$    & 1.5   & -19.063 & -18.685 & -23.769 & -18.691 & -27.156 & -22.737 & -29.298 & -24.03 \\
 10$^{9}$ & 2     & -14.208 & -13.825 & -17.813 & -13.836 & -19.979 & -16.773 & -21.461 & -17.72 \\
 10$^{9}$    & 3     & -9.189 & -8.802 & -11.653 & -8.830 & -12.601 & -10.662 & -13.43 & -11.236 \\
 10$^{9}$    & 5     & -4.920 & -4.531 & -6.231 & -4.587 & -6.384 & -5.507 & -6.701 & -5.738 \\
 10$^{9}$  & 10    & -1.101 & -0.835 & -1.295 & -0.883 & -1.202 & -1.019 & -1.192 & -0.968 \\
 10$^{9}$   & 15    & 0.695 & 0.807 & 0.689 & 0.782 & 0.784 & 0.821 & 0.857 & 0.934 \\
 10$^{9}$   & 20    & 1.770  & 1.801 & 1.798 & 1.788 & 1.881 & 1.856 & 1.966 & 1.987 \\
10$^{9}$   & 25    & 2.501 & 2.487 & 2.539 & 2.482 & 2.613 & 2.553 & 2.699 & 2.690 \\
 10$^{9}$  & 30    & 3.062 & 3.021 & 3.104 & 3.022 & 3.172 & 3.089 & 3.256 & 3.229 \\
          &       &       &       &       &       &       &       &       &  \\
10$^{11}$  & 0.4   & 4.649 & 4.850  & 2.556 & 4.007 & 2.174 & 2.499 & 1.576 & 2.052 \\
 10$^{11}$   & 0.7   & 4.650  & 4.887 & 2.598 & 4.351 & 2.223 & 2.537 & 1.664 & 2.147 \\
  10$^{11}$   & 1     & 4.650  & 4.910  & 2.614 & 4.526 & 2.240  & 2.547 & 1.732 & 2.174 \\
10$^{11}$   & 1.5   & 4.651 & 4.937 & 2.620  & 4.682 & 2.247 & 2.551 & 1.798 & 2.178 \\
 10$^{11}$   & 2     & 4.653 & 4.959 & 2.624 & 4.765 & 2.247 & 2.594 & 1.834 & 2.180 \\
 10$^{11}$  & 3     & 4.661 & 4.995 & 2.714 & 4.848 & 2.273 & 3.053 & 1.878 & 2.486 \\
 10$^{11}$   & 5     & 4.684 & 5.045 & 3.379 & 4.920  & 3.149 & 3.949 & 2.773 & 3.653 \\
 10$^{11}$  & 10    & 4.951 & 5.211 & 4.738 & 5.134 & 4.801 & 4.970  & 4.783 & 4.993 \\
 10$^{11}$   & 15    & 5.387 & 5.498 & 5.371 & 5.456 & 5.449 & 5.478 & 5.505 & 5.574 \\
 10$^{11}$  & 20    & 5.704 & 5.735 & 5.726 & 5.711 & 5.798 & 5.768 & 5.872 & 5.888 \\
 10$^{11}$  & 25    & 5.919 & 5.905 & 5.952 & 5.892 & 6.019 & 5.955 & 6.096 & 6.084 \\
 10$^{11}$  & 30    & 6.073 & 6.032 & 6.111 & 6.027 & 6.174 & 6.089 & 6.252 & 6.223 \\
 \bottomrule
\end{tabular}}}
\end{table}

\begin{table}[tp]
\caption{\small The same as table \ref{table3} but for $\beta^{-}$ rates ($\lambda_{\beta^{-}}$) due to unique-first forbidden (U1F) transitions.
}\label{table5} \centering {\resizebox{!}{11.5cm}{%
\begin{tabular}{cccccccccccc}
\toprule \multirow{2}{*}{$\rho$Y$_{e}$} &
\multirow{2}{*}{T$_{9}$} & \multicolumn{8}{c}{$\lambda_{\beta^{-}} [s^{-1}]$ (U1F)} \\
\cmidrule{3-4}  \cmidrule{5-6}  \cmidrule{7-8} \cmidrule{9-10}& &
\multicolumn{1}{c}{$^{101}$Y} &
\multicolumn{1}{c}{$^{102}$Y} &
\multicolumn{1}{c}{$^{103}$Y} &
\multicolumn{1}{c}{$^{104}$Y} &
\multicolumn{1}{c}{$^{105}$Y} &
\multicolumn{1}{c}{$^{106}$Y} &
\multicolumn{1}{c}{$^{107}$Y} &
\multicolumn{1}{c}{$^{108}$Y} \\
\midrule
 10$^{3}$  & 0.4   & -0.867 & -1.017 & -0.013 & -0.844 & 0.468 & 0.277 & 1.256 & 1.106 \\
 10$^{3}$ & 0.7   & -0.874 & -0.799 & -0.026 & -0.723 & 0.442 & 0.361 & 1.254 & 1.142 \\
 10$^{3}$ & 1     & -0.877 & -0.696 & -0.034 & -0.591 & 0.430  & 0.479 & 1.253 & 1.235 \\
10$^{3}$  & 1.5   & -0.880 & -0.610 & -0.040 & -0.377 & 0.424 & 0.658 & 1.256 & 1.400 \\
10$^{3}$  & 2     & -0.879 & -0.550 & -0.043 & -0.182 & 0.427 & 0.781 & 1.264 & 1.515 \\
 10$^{3}$  & 3     & -0.872 & -0.438 & -0.044 & 0.109 & 0.443 & 0.923 & 1.286 & 1.643 \\
10$^{3}$ & 5     & -0.850 & -0.237 & -0.036 & 0.411 & 0.484 & 1.055 & 1.331 & 1.750 \\
10$^{3}$  & 10    & -0.592 & 0.112 & 0.234 & 0.731 & 0.826 & 1.249 & 1.655 & 1.881 \\
10$^{3}$   & 15    & -0.183 & 0.396 & 0.665 & 0.990  & 1.271 & 1.467 & 2.085 & 2.064 \\
10$^{3}$  & 20    & 0.095 & 0.604 & 0.951 & 1.188 & 1.558 & 1.647 & 2.354 & 2.227 \\
10$^{3}$  & 25    & 0.266 & 0.738 & 1.125 & 1.317 & 1.731 & 1.768 & 2.513 & 2.339 \\
10$^{3}$  & 30    & 0.375 & 0.824 & 1.234 & 1.401 & 1.841 & 1.846 & 2.612 & 2.412 \\
          &       &       &       &       &       &       &       &       &  \\
 10$^{6}$   & 0.4   & -0.870 & -1.019 & -0.014 & -0.846 & 0.468 & 0.276 & 1.256 & 1.106 \\
  10$^{6}$  & 0.7   & -0.877 & -0.801 & -0.027 & -0.725 & 0.442 & 0.360  & 1.254 & 1.141 \\
   10$^{6}$  & 1     & -0.881 & -0.698 & -0.035 & -0.593 & 0.430  & 0.478 & 1.253 & 1.234 \\
  10$^{6}$ & 1.5   & -0.883 & -0.612 & -0.041 & -0.379 & 0.424 & 0.657 & 1.256 & 1.399 \\
 10$^{6}$ & 2     & -0.882 & -0.552 & -0.044 & -0.183 & 0.427 & 0.780  & 1.264 & 1.515 \\
  10$^{6}$ & 3     & -0.874 & -0.439 & -0.045 & 0.108 & 0.442 & 0.922 & 1.286 & 1.643 \\
  10$^{6}$ & 5     & -0.851 & -0.238 & -0.037 & 0.410  & 0.483 & 1.055 & 1.331 & 1.749 \\
 10$^{6}$  & 10    & -0.593 & 0.112 & 0.234 & 0.730  & 0.826 & 1.248 & 1.655 & 1.881 \\
  10$^{6}$ & 15    & -0.183 & 0.396 & 0.664 & 0.990  & 1.270  & 1.467 & 2.084 & 2.064 \\
  10$^{6}$& 20    & 0.095 & 0.604 & 0.951 & 1.188 & 1.557 & 1.647 & 2.354 & 2.227 \\
  10$^{6}$ & 25    & 0.266 & 0.738 & 1.124 & 1.317 & 1.731 & 1.768 & 2.513 & 2.339 \\
  10$^{6}$  & 30    & 0.375 & 0.824 & 1.234 & 1.401 & 1.841 & 1.846 & 2.612 & 2.412 \\
          &       &       &       &       &       &       &       &       &  \\
10$^{9}$   & 0.4   & -5.288 & -11.879 & -1.318 & -3.78 & -0.31 & -1.593 & 0.873 & 0.333 \\
10$^{9}$  & 0.7   & -5.122 & -8.115 & -1.327 & -3.048 & -0.349 & -1.026 & 0.860  & 0.412 \\
10$^{9}$ & 1     & -4.864 & -6.457 & -1.327 & -2.697 & -0.365 & -0.657 & 0.848 & 0.589 \\
 10$^{9}$ & 1.5   & -4.42 & -4.99 & -1.315 & -2.263 & -0.367 & -0.307 & 0.840  & 0.844 \\
10$^{9}$ & 2     & -4.039 & -4.116 & -1.293 & -1.897 & -0.354 & -0.118 & 0.846 & 1.000 \\
10$^{9}$  & 3     & -3.467 & -3.050 & -1.230 & -1.372 & -0.309 & 0.087 & 0.874 & 1.165 \\
 10$^{9}$  & 5     & -2.741 & -1.945 & -1.065 & -0.76 & -0.193 & 0.314 & 0.948 & 1.311 \\
 10$^{9}$  & 10    & -1.606 & -0.748 & -0.433 & 0.040  & 0.350  & 0.743 & 1.359 & 1.545 \\
10$^{9}$ & 15    & -0.762 & -0.102 & 0.252 & 0.573 & 0.955 & 1.141 & 1.876 & 1.828 \\
10$^{9}$   & 20    & -0.241 & 0.310  & 0.699 & 0.935 & 1.356 & 1.442 & 2.212 & 2.070 \\
 10$^{9}$   & 25    & 0.069 & 0.562 & 0.970  & 1.163 & 1.604 & 1.639 & 2.419 & 2.236 \\
10$^{9}$   & 30    & 0.254 & 0.714 & 1.136 & 1.303 & 1.758 & 1.762 & 2.548 & 2.343 \\
          &       &       &       &       &       &       &       &       &  \\
 10$^{11}$  & 0.4   & -100  & -100  & -100  & -100  & -100  & -100  & -100  & -100 \\
 10$^{11}$  & 0.7   & -100  & -100  & -100  & -100  & -100  & -100  & -100  & -100 \\
  10$^{11}$ & 1     & -98.931 & -100  & -91.747 & -95.720 & -86.238 & -85.638 & -76.024 & -77.069 \\
  10$^{11}$ & 1.5   & -67.319 & -68.021 & -62.118 & -64.341 & -58.267 & -57.573 & -51.25 & -51.652 \\
  10$^{11}$ & 2     & -51.260 & -51.395 & -47.135 & -48.501 & -44.152 & -43.458 & -38.751 & -38.855 \\
  10$^{11}$& 3     & -34.977 & -34.579 & -31.982 & -32.501 & -29.862 & -29.217 & -26.083 & -25.935 \\
 10$^{11}$ & 5     & -21.683 & -20.884 & -19.625 & -19.487 & -18.169 & -17.588 & -15.678 & -15.401 \\
 10$^{11}$  & 10    & -11.149 & -10.267 & -9.777 & -9.368 & -8.753 & -8.376 & -7.213 & -7.102 \\
  10$^{11}$  & 15    & -7.178 & -6.483 & -6.006 & -5.719 & -5.142 & -4.970 & -3.894 & -4.006 \\
   10$^{11}$  & 20    & -5.076 & -4.485 & -3.999 & -3.782 & -3.216 & -3.139 & -2.119 & -2.316 \\
   10$^{11}$  & 25    & -3.787 & -3.254 & -2.766 & -2.586 & -2.029 & -2.000    & -1.023 & -1.253 \\
  10$^{11}$  & 30    & -2.919 & -2.422 & -1.933 & -1.775 & -1.225 & -1.225 & -0.280 & -0.524 \\
 \bottomrule
\end{tabular}}}
\end{table}

\begin{table}[tp]
\caption{\small The same as table \ref{table3} but for EC rates ($\lambda_{EC}$)
due to unique-first forbidden (U1F) transitions.}\label{table6} \centering {\resizebox{!}{11.5cm}{%
\begin{tabular}{cccccccccccc}
\toprule \multirow{2}{*}{$\rho$Y$_{e}$} &
\multirow{2}{*}{T$_{9}$} & \multicolumn{8}{c}{$\lambda_{EC} [s^{-1}]$ (U1F)} \\
\cmidrule{3-4}  \cmidrule{5-6}  \cmidrule{7-8} \cmidrule{9-10}& &
\multicolumn{1}{c}{$^{101}$Y} &
\multicolumn{1}{c}{$^{102}$Y} &
\multicolumn{1}{c}{$^{103}$Y} &
\multicolumn{1}{c}{$^{104}$Y} &
\multicolumn{1}{c}{$^{105}$Y} &
\multicolumn{1}{c}{$^{106}$Y} &
\multicolumn{1}{c}{$^{107}$Y} &
\multicolumn{1}{c}{$^{108}$Y} \\
\midrule
 10$^{3}$  & 0.4   & -100  & -100  & -100  & -100  & -100  & -100  & -100  & -100 \\
 10$^{3}$  & 0.7   & -75.177 & -74.881 & -85.011 & -74.922 & -93.999 & -83.800 & -99.237 & -86.747 \\
  10$^{3}$  & 1     & -53.132 & -52.920 & -60.097 & -52.828 & -66.018 & -58.973 & -69.551 & -61.019 \\
  10$^{3}$   & 1.5   & -35.101 & -34.943 & -39.846 & -34.753 & -43.332 & -38.787 & -45.539 & -40.118 \\
 10$^{3}$   & 2     & -25.872 & -25.721 & -29.506 & -25.490 & -31.746 & -28.487 & -33.287 & -29.458 \\
10$^{3}$   & 3     & -16.459 & -16.272 & -18.942 & -16.023 & -19.937 & -18.014 & -20.806 & -18.626 \\
10$^{3}$   & 5     & -8.632 & -8.355 & -9.950 & -8.140 & -10.131 & -9.342 & -10.471 & -9.661 \\
10$^{3}$   & 10    & -2.013 & -1.803 & -2.217 & -1.666 & -2.121 & -2.165 & -2.132 & -2.225 \\
10$^{3}$  & 15    & 0.837 & 0.822 & 0.815 & 0.930  & 0.935 & 0.700   & 0.974 & 0.706 \\
10$^{3}$    & 20    & 2.513 & 2.374 & 2.518 & 2.471 & 2.641 & 2.337 & 2.682 & 2.359 \\
10$^{3}$   & 25    & 3.633 & 3.428 & 3.642 & 3.522 & 3.764 & 3.425 & 3.800   & 3.451 \\
 10$^{3}$  & 30    & 4.449 & 4.205 & 4.457 & 4.296 & 4.578 & 4.217 & 4.607 & 4.244 \\
          &       &       &       &       &       &       &       &       &  \\
10$^{6}$  & 0.4   & -100  & -100  & -100  & -100  & -100  & -100  & -100  & -100 \\
10$^{6}$   & 0.7   & -71.554 & -71.259 & -81.389 & -71.300 & -90.377 & -80.177 & -95.615 & -83.125 \\
 10$^{6}$  & 1     & -49.975 & -49.763 & -56.940 & -49.671 & -62.861 & -55.816 & -66.393 & -57.862 \\
10$^{6}$   & 1.5   & -33.089 & -32.930 & -37.833 & -32.740 & -41.319 & -36.775 & -43.527 & -38.105 \\
10$^{6}$   & 2     & -24.584 & -24.433 & -28.218 & -24.202 & -30.459 & -27.200 & -31.999 & -28.170 \\
10$^{6}$  & 3     & -15.957 & -15.770 & -18.440 & -15.522 & -19.435 & -17.512 & -20.305 & -18.124 \\
10$^{6}$   & 5     & -8.540 & -8.262 & -9.858 & -8.048 & -10.039 & -9.250 & -10.379 & -9.569 \\
10$^{6}$  & 10    & -2.003 & -1.793 & -2.207 & -1.656 & -2.111 & -2.155 & -2.122 & -2.215 \\
10$^{6}$  & 15    & 0.840  & 0.825 & 0.818 & 0.933 & 0.938 & 0.703 & 0.977 & 0.709 \\
10$^{6}$  & 20    & 2.514 & 2.375 & 2.519 & 2.473 & 2.642 & 2.338 & 2.684 & 2.360 \\
10$^{6}$  & 25    & 3.634 & 3.429 & 3.643 & 3.522 & 3.765 & 3.426 & 3.800   & 3.452 \\
10$^{6}$  & 30    & 4.450  & 4.205 & 4.458 & 4.297 & 4.579 & 4.217 & 4.608 & 4.244 \\
          &       &       &       &       &       &       &       &       &  \\
10$^{9}$   & 0.4   & -68.965 & -68.458 & -86.017 & -68.86 & -100  & -84.553 & -100  & -89.694 \\
10$^{9}$   & 0.7   & -39.312 & -39.016 & -49.147 & -39.057 & -58.135 & -47.935 & -63.372 & -50.883 \\
 10$^{9}$ & 1     & -27.277 & -27.064 & -34.242 & -26.973 & -40.163 & -33.118 & -43.695 & -35.164 \\
10$^{9}$   & 1.5   & -17.747 & -17.588 & -22.491 & -17.398 & -25.977 & -21.433 & -28.185 & -22.763 \\
10$^{9}$  & 2     & -12.868 & -12.717 & -16.502 & -12.486 & -18.743 & -15.484 & -20.284 & -16.454 \\
10$^{9}$  & 3     & -7.828 & -7.641 & -10.311 & -7.392 & -11.306 & -9.383 & -12.175 & -9.995 \\
10$^{9}$  & 5     & -3.530 & -3.252 & -4.847 & -3.038 & -5.029 & -4.239 & -5.368 & -4.558 \\
 10$^{9}$   & 10    & 0.360  & 0.569 & 0.156 & 0.707 & 0.252 & 0.208 & 0.241 & 0.148 \\
10$^{9}$   & 15    & 2.225 & 2.21  & 2.203 & 2.318 & 2.323 & 2.088 & 2.362 & 2.094 \\
10$^{9}$  & 20    & 3.367 & 3.228 & 3.372 & 3.325 & 3.495 & 3.191 & 3.536 & 3.213 \\
10$^{9}$  & 25    & 4.161 & 3.956 & 4.170  & 4.050  & 4.292 & 3.953 & 4.328 & 3.979 \\
 10$^{9}$   & 30    & 4.780  & 4.536 & 4.788 & 4.628 & 4.909 & 4.548 & 4.938 & 4.575 \\
          &       &       &       &       &       &       &       &       &  \\
 10$^{11}$   & 0.4   & 6.344 & 6.358 & 4.373 & 5.01  & 4.001 & 2.5   & 3.733 & 2.093 \\
 10$^{11}$ & 0.7   & 6.344 & 6.358 & 4.364 & 5.780  & 3.990  & 2.699 & 3.761 & 2.127 \\
 10$^{11}$  & 1     & 6.344 & 6.360  & 4.356 & 6.079 & 3.980 & 2.860  & 3.785 & 2.222 \\
 10$^{11}$ & 1.5   & 6.345 & 6.372 & 4.344 & 6.319 & 3.966 & 3.071 & 3.815 & 2.408 \\
  10$^{11}$  & 2     & 6.348 & 6.400   & 4.342 & 6.452 & 3.961 & 3.511 & 3.836 & 2.683 \\
  10$^{11}$ & 3     & 6.357 & 6.478 & 4.431 & 6.612 & 3.986 & 4.531 & 3.871 & 3.822 \\
  10$^{11}$ & 5     & 6.384 & 6.623 & 5.083 & 6.773 & 4.824 & 5.513 & 4.488 & 5.133 \\
  10$^{11}$  & 10    & 6.675 & 6.865 & 6.445 & 6.976 & 6.511 & 6.451 & 6.475 & 6.367 \\
  10$^{11}$  & 15    & 7.134 & 7.108 & 7.096 & 7.200   & 7.200   & 6.956 & 7.225 & 6.948 \\
   10$^{11}$ & 20    & 7.480  & 7.332 & 7.475 & 7.42  & 7.587 & 7.277 & 7.619 & 7.290 \\
 10$^{11}$ & 25    & 7.726 & 7.515 & 7.728 & 7.602 & 7.843 & 7.499 & 7.872 & 7.520 \\
   10$^{11}$ & 30    & 7.914 & 7.664 & 7.917 & 7.751 & 8.033 & 7.667 & 8.057 & 7.691 \\
 \bottomrule
\end{tabular}}}
\end{table}
In tables~\ref{table3} and~\ref{table4}, the results of our
calculated $\beta^{-}$ rates ($\lambda_{\beta^{-}}$) and EC rates
($\lambda_{EC}$) for allowed GT transitions have been shown,
respectively. Similarly, tables~\ref{table4} and~\ref{table5}
present $\lambda_{\beta^{-}}$ and $\lambda_{EC}$ rates for U1F
transitions, respectively. As mentioned earlier, our calculations
are performed for large stellar temperature and density domains.
Temperature (T$_{9}$) and densities ($\rho$Y$_{e}$) are given in
units of $10^{9}$\;K and gcm$^{-3}$, respectively. This domain of
temperature also covers astrophysically important r-process range
from T$_{9}$ = 1 to 3. We have shown the rates at four different
stellar density values, which correspond to low
($\rho$Y$_{e}$\;=\;10$^{3}$), medium ($\rho$Y$_{e}$\;=\;10$^{6}$)
and high ($\rho$Y$_{e}$\;=\;10$^{9}$,\;10$^{11}$) density regions.
The first and second columns of each table show the selected density
and temperature values, respectively. The remaining columns show the
rate values of $^{101-108}$Y isotopes computed in this work. Results
presented in table \ref{table3} depict that with the increase in
stellar core density, allowed $\lambda_{\beta^{-}}$ values get
smaller. The rate of decrement in the values of
$\lambda_{\beta^{-}}$ increases as one goes towards high density
regions. EC and $\beta^{-}$-decay rates largely depend on
the available phase space which shows considerable expansion
(contraction) with temperature (density). The contraction of the
available phase space for electrons with increase in density is due
to Pauli blocking. This is one of the key reasons for the reduction
in rates in $\beta^{-}$ direction as the density goes high. Both EC
and $\beta^{-}$-decay rates in each density region show an
increasing trend with the rise of temperature due to phase space
expansion. Additionally, with increasing temperature, the occupation
probabilities of parent excited states increase. In the result,
partial rates of parent excited states substantially contribute in
increasing the values of total rates. Therefore, increment in
$\lambda_{\beta^{-}}$ is noted as core temperature goes up.  The
increment is large in higher ($\rho$Y$_{e}$\;=\;10$^{11}$) density
region as compared to lower ($\rho$Y$_{e}$\;=\;10$^{3}$) and medium
($\rho$Y$_{e}$\;=\;10$^{6}$) densities, especially in T$_{9} \leq
10$. However, as the density of the stellar core increases, the
Fermi energy (E$_{f}$) of the electrons raises which causes an
enhancement in electron capture rates. Therefore, in contrast to
$\lambda_{\beta^{-}}$, with increasing density, $\lambda_{EC}$
values get higher in magnitude. The rate of increment in
$\lambda_{EC}$ is larger in higher density region. Like
$\lambda_{\beta^{-}}$, EC rates also show increasing trend with
temperature and become many orders of magnitude larger in
$\rho$Y$_{e}$\;$\geq$\;10$^{9}$ region, at T$_{9}$ = 25, 30, as
compared to lower temperatures.

Our calculated values of U1F rates in both $\beta^{-}$ and EC directions exhibit a trend similar
to the corresponding allowed rates. Table \ref{table5} shows the change in forbidden
$\lambda_{\beta^{-}}$ values with increase in stellar temperature and density.
It is to be noticed that with rising density, $\beta^{-}$ rates decrease and this decrement  approaches to many orders of magnitude in the high density
($\rho$Y$_{e}$\;=\;10$^{11}$) region. However, these rates increase with rising stellar temperature.
The forbidden $\lambda_{EC}$ increase with both rising temperature and density
(table \ref{table6}). Like allowed EC, many orders of magnitude increment can be
noticed in forbidden EC rates at high
density and temperature.

Lastly, we move towards the comparison of our calculated allowed and
U1F rates for both $\beta^{-}$-decay and EC processes. In case of
$\lambda_{\beta^{-}}$,  calculated allowed rates in general are
greater than U1F rates, for each value of density, as can be seen
from tables~\ref{table3} and~\ref{table5}. On the contrary,
tables~\ref{table4} and~\ref{table6} show that EC U1F rates are
larger than allowed rates over all density and temperature domains.
The reason of this enhancement in U1F rates in EC direction is due
to the phase space amplification because of U1F transitions. In
stellar environments, the phase space integrals of U1F and allowed
GT transitions compete well with each other. However, under specific
stellar conditions, U1F phase space override the allowed one.

\section{Conclusions}
\label{sec:conclusions}

In this study, we have calculated several $\beta$-decay properties of eight Y isotopes having A=101-108. The calculations of
$\beta$-decay half-lives under terrestrial conditions, charge-changing strength distributions, weak rates
both in $\beta^{-}$ and EC directions under stellar conditions and beta-delayed neutron emission probabilities
have been performed by employing the pn-QRPA model. Our work shows a decent agreement of computed T$_{\beta_{1/2}}$
with the experimentally measured values, once the U1F transitions are included in the calculations. The ratios between
our results and experimental measurements are within a factor of 1.3. A comparison between our allowed and
U1F charge-changing strength distributions shows reduction in half-life values which becomes quite significant
in cases of $^{107}$Y ($\sim 67\%$) and $^{108}$Y ($\sim 42\%$).

Our calculated allowed and U1F $\beta^{-}$-decay rates decrease (increase) with increasing stellar density (temperature),
mainly in the high density regions. On the contrary, EC rates due to both of the allowed and U1F transitions rise with
increasing temperature and density. However, the increment in rates with density (temperature) is more visible in high (medium) density region.
The comparison between our allowed and U1F rates, in general, shows an opposite trend in case of $\beta^{-}$ and EC transitions.
For $\beta^{-}$ transitions, allowed rates take lead on U1F rates. In contrast, U1F rates are larger in magnitude than allowed
rates in case of EC transitions. The simulators may find our results interesting to be used as an input for the simulation and modeling of r-process.


\section*{References}
\begin{thebibliography}{99}
\bibitem{Hoy46} Hoyle F 1946 {\it MNRAS} \textbf{106} 343-383
\bibitem{Hoy56} Hoyle F, Fowler W A, Burbidge G R and Burbidge 1956 {\it E. M.: Science} \textbf{124} 611
\bibitem{Cam57} Cameron A G W 1957 {\it PASP} \textbf{69} 201
\bibitem{Bur57} Burbidge E M, Burbidge G R, Fowler W A and Hoyle F 1957 {\it Rev. Mod. Phys.} \textbf{29} 547
\bibitem{Cyb16} Cyburt R H, Fields B D, Olive K A and Yeh T H 2016 {\it Rev. Mod. Phys.} \textbf{88} 015004
\bibitem{Joh19} Johnson J A 2019 {\it Science} \textbf{363} 474-478
\bibitem{Kap89} Kappeler F, Beer H and Wisshak K 1989 {\it Rep. Prog. Phys.} \textbf{52} 945
\bibitem{Kap11} Kappeler F, Gallino R, Bisterzo S and Aoki W 2011 {\it Rev. Mod. Phys.} \textbf{83} 157
\bibitem{Sat74} Sato K 1974 {\it Prog. Theor. Phys.} \textbf{51} 726
\bibitem{Cow91} Cowan J J, Thielemann F-K and Truran J W 1991 {\it Phys. Rep.} \textbf{208} 267
\bibitem{Amo07} Arnould M, Goriely S and Takahashi K 2007 {\it Phys. Rep.} \textbf{450} 97
\bibitem{Horo19} Horowitz C J, Arcones A, C\^{o}t\'{e} B and Dillmann I {\it et al} 2019 {\it J. Phys. G: Nucl. Part. Phys.} \textbf{46} 083001
\bibitem{Cow21} Cowan J J, Sneden C, Lawler J E, Aprahamian A, Wiescher M, Langanke K, \etal 2021 {\it Rev. Mod. Phys.} \textbf{93} 015002
\bibitem{Wat19} Watson D, Hansen C J, Selsing J, Koch A, Malesani D B, Andersen A C, \etal 2019 {\it Nature} \textbf{574} 497
\bibitem{Kra93} Kratz K-L, Bitouzet J-P, Thielemann F-K, M\"{o}ller P, Pfeiffer B 1993 {\it Ap. J.} \textbf{403} 216
\bibitem{Arc11} Arcones A and Mart´ınez-Pinedo G 2011 {\it Phys. Rev. C} \textbf{83} 045809
\bibitem{Mum16} Mumpower M R, Surman R, McLaughlin G C and Aprahamian A 2016 {\it Progress in Particle and Nuclear Physics} \textbf{86} 86
\bibitem{Kurt09} Kurtukian-Nieto T \etal 2009 {\it Nucl. Phys. A}  \textbf{827} 587
\bibitem{Nish11} Nishimura S \etal 2011 {\it Phys. Rev. Lett.} \textbf{106} 052502
\bibitem{Hal21} Hall O, Davinson T, Estrade A and Liu J {\it et al} 2021 {\it Phys. Lett. B} \textbf{816} 136266
\bibitem{Klpa1984} Klapdor-Kleingrothaus H V, Metzinger J, Oda T 1984 {\it At. Data Nucl. Data Tables} \textbf{31} 81
\bibitem{Sta89} Staudt A, Bender E, Muto K and Klapdor-Kleingrothaus H V 1989 {\it Z. Phys. A — Atomic Nuclei} ${\bf 334}$ 47
\bibitem{Sta90} Staudt A, Bender E, Muto K and Klapdor-Kleingrothaus H V 1990 {\it At. Data Nucl. Data Tables} ${\bf 44}$ 132
\bibitem{Hir93} Hirsch M, Staudt A, Muto K, Klapdor-Kleingrothaus H V 1993 {\it At. Data Nucl. Data Tables} \textbf{53} 165
\bibitem{Nabi04} Nabi J-U and Klapdor-Kleingrothaus H V 2004 {\it At. Data Nucl. Data Tables} ${\bf 88}$ 237
\bibitem{Nabi19} Nabi J-U, Ullah A, Shah S A A, Daraz G and Ahmad M 2019 {\it Act. Phys. Pol. B} \textbf{50} 1523
\bibitem{Nabi2021} Nabi J-U, Bayram T, Daraz G, Kabir A and \c{S}ent\"{u}rk \c{S} 2021 {\it Nucl. Phys. A} \textbf{1015} 122278
\bibitem{Nabi21} Nabi J-U, \c{C}akmak N, Ullah A and Khan A U 2021 {\it Phys. Scr.} \textbf{96} 115303
\bibitem{Nabi2020} Nabi, J-U, B\"{o}y\"{u}kata M, Ullah A, and Riaz M 2020 {\it Nucl. Phys. A} \textbf{1002} 121985
\bibitem{Fuller} Fuller G M, Fowler W A, Newman M J 1980 {\it Astrophys. J. Suppl. Ser.} ${\bf 42}$ 447; 1982a {\it Astrophys. J. Suppl. Ser.} ${\bf 48}$ 279; 1982b {\it Astrophys. J.} ${\bf 252}$ 715; 1985 {\it Astrophys. J.} ${\bf 293}$ 1
\bibitem{Lang00} Langanke K, Mart\'{i}nez-Pinedo G 2000 {\it Nucl. Phys. A} ${\bf673}$ 481
\bibitem{LangMar03} Langanke K, Mart\'{i}nez-Pinedo G, Sampaio J M, Dean D J, Hix W R, Messer O E B, Mezzacappa A, Liebendörfer M, Janka H-Th, and Rampp M2003 {\it Phys. Rev. Lett.} ${\bf90}$ 241102
\bibitem{Sar10} Sarriguren P and Pereira J 2010 {\it Phys. Rev. C} \textbf{81} 064314
\bibitem{Sar14} Sarriguren P, Algora A and Pereira J 2014 {\it Phys. Rev. C} \textbf{89} 034311
\bibitem{Lor15} Lorusso G \etal 2015 {\it Phys. Rev. Lett.} \textbf{114}(19) 192501
\bibitem{Zhi18} Li Z, Zhou Y, Li X, Wang Y, Guo B, Nan D and Liu W 2018 {\it EPJ Web Conf.} \textbf{178} 04006
\bibitem{Hom96} Homma H, Bender E, Hirsch M, Muto K, Klapdor-Kleingrothaus H V and Oda T 1996 {\it Phys. Rev. C} ${\bf 54}$ 2972
\bibitem{Mol03} M\"{o}ller P, Pfeiffer B, Kratz K-L 2003 {\it Phys. Rev. C} \textbf{67} 055802
\bibitem{Nil55} Nilsson S G 1955 {\it Mat. Fys. Medd. Dan. Vid. Selsk} \textbf{29} no. 16
\bibitem{Nabi99} Nabi J-U and Klapdor-Kleingrothaus H V 1999 {\it Eur. Phys. J. A} ${\bf 5}$ 337.
\bibitem{Aud17} Audi G, Kondev F G, Wang M, Huang W J, and Naimi S 2017 {\it Chin. Phys. C} \textbf{41} 030001
\bibitem{Mol81} M\"{o}ller P and Nix J R 1981 {\it At. Data Nucl. Data Tables} \textbf{26} 165
\bibitem{Boh69} Bohr A and Mottelson B R 1969 {\it Nuclear Structure Vol. 1} (Benjamin, New York)
\bibitem{Muto92} Muto K, Bender E, Oda T and Klapdor-Kleingrothaus H V 1992 {\it Z. Phys. A} ${\bf 341}$ 407
\bibitem{Nak10} Nakamura K 2010 {\it Particle Data Group J. Phys. G: Nucl. and Part. Phys.} ${\bf37 (7A)}$ 075021
\bibitem{Har09} Hardy J C, Towner I C 2009 {\it Phys. Rev. C} ${\bf79 (5)}$ 055502
\bibitem{Gove71} Gove N B and Martin M J 1971 {\it At. Data Nucl. Data Tables} $\bf{10}$ 205
\bibitem{Pfe02} Pfeiffer B, Kratz K-L, M\"{o}ller P 2002 {\it Prog. Nucl. Energy} \textbf{41} 39
\bibitem{Krum84} Krumlinde J and M\"{o}ller P 1984 \textit{Nucl. Phys. A} \textbf{417} 419
\bibitem{Mol90} M\"{o}ller P and Randrup J 1990 {\it Nucl. Phys. A} \textbf{514} 1
\bibitem{Mol92} M\"{o}ller P {\it et al} 1992 {\it Nucl. Phys. A} \textbf{536}, 61
\bibitem{Kratz73} Kratz K-L and Herrmann G 1973 {\it Z. Physik.} \textbf{263} 435
\bibitem{Pfe00} Pfeiffer Kratz K-L and M\"{o}ller P 2000 {\it Internal Report, Inst. f\"{u}r Kernchemie, Univ. Mainz} URL: www.kernchemie.uni-mainz.de/pfeiffer/khf/
\end{thebibliography}
\end{document}